\documentclass[10pt]{article}

\usepackage{amsmath,amssymb,dsfont,slashed,color,hyperref}
\usepackage{authblk}
\usepackage{enumerate}
\usepackage[toc,page]{appendix}
\numberwithin{equation}{section}

\definecolor{armygreen}{rgb}{0.29, 0.33, 0.13}

\newcommand{\half}{\frac{1}{2}}

\newcommand{\JJ}{\mathds{J}}
\newcommand{\KK}{\mathds{K}}
\newcommand{\DD}{\mathds{D}}
\newcommand{\TT}{\mathds{T}}
\newcommand{\ZZ}{\mathds{Z}}

\newcommand{\QQ}{\mathds{Q}}
\newcommand{\QQb}{\overline{\mathds{Q}}}

\newcommand{\AAA}{\mathds{A}}
\newcommand{\FF}{\mathds{F}}
\newcommand{\EE}{\mathds{E}}
\newcommand{\OO}{\mathds{O}}

\newcommand{\XX}{\mathds{X}}
\newcommand{\calF}{\mathcal{F}}
\newcommand{\calX}{\mathcal{X}}
\newcommand{\ocalX}{\overline{\mathcal{X}}}
\newcommand{\calG}{\mathcal{G}}
\newcommand{\calH}{\mathcal{H}}
\newcommand{\calL}{\mathcal{L}}
\newcommand{\calR}{\mathcal{R}}
\newcommand{\calO}{\mathcal{O}}
\newcommand{\se}{\slashed{e}}
\newcommand{\sT}{\slashed{T}}
\newcommand{\sD}{\slashed{D}}

\newcommand{\Om}{{\Omega^-}}

\newcommand{\lD}{\overleftarrow{D}}

\newcommand{\lsD}{\overleftarrow{\sD}}
\newcommand{\psibar}{{\overline{\psi}}}
\newcommand{\epsilonbar}{{\overline{\epsilon}}}
\newcommand{\tgamma}{\tilde{\gamma}}
\newcommand{\gf}{\gamma_5 }
\newcommand{\dV}{|e|d^4x}

\newcommand{\MP}{{M_\text{P}}}
\newcommand{\GN}{{G_\text{N}}}

\newcommand{\uantof}{Departamento de Física, Universidad de Antofagasta, Aptdo. 02800, Chile}

\newcommand{\cecs}{Centro de Estudios Cient\'{\i}ficos (CECs), Arturo Prat 514, Valdivia, Chile}

\begin{document}

\title{Gauging the superconformal group with a graded dual operator}
\author[1]{P. D. Alvarez \thanks{E-mail: \href{mailto:pedro.alvarez@uantof.cl}{\nolinkurl{pedro.alvarez@uantof.cl}}}}

\author[1]{R. A. Chavez \thanks{E-mail: \href{mailto:rafael.chavez.linares@uantof.cl}{\nolinkurl{rafael.chavez.linares@uantof.cl}}}}
\affil[1]{\uantof}

\author[2]{J. Zanelli \thanks{E-mail: \href{mailto:z@cecs.cl}{\nolinkurl{z@cecs.cl}}}}
\affil[2]{\cecs}

\maketitle

\begin{abstract}
 Based on the superconformal algebra we construct a dual operator that introduces a grading among bosonic generators independent of the boson/fermion grading of the superalgebra. This dual operator allows us to construct an action that is gauge invariant under the grading even bosonic generators. We provide a self-dual notion based on the dual operator. We use the definition of the dual operator to contruct a model with gauge invariance $SO(1,3)\times SU(N) \times U(1) \subset SU(2,2|N)$. The choice of a graded dual operator allows us to overcome technical difficulties of $U(N)$ unified theories based on the superconformal group. The gravity action reproduces the Einstein-Hilbert action in certain sector of the theory. The definition of the dual operator allows us to include fermionic matter in the gauge connection in a geometric manner. We give a summary with possible phenomenological parameters that are included in the model.
\end{abstract}

\tableofcontents

\section{Introduction}\label{intro}

In the late seventies there were attempts to construct unified theories with matter \cite{Ferrara:1977ij} using the superconformal algebra $su(2,2|N)$. Such algebra is the symmetry algebra of $N$-extended conformal supergravity in four dimensions, which contains the conformal algebra $o(4,2) \sim su(2,2)$, and $u(N)$ as subalgebras \cite{Fradkin:1985am}. The conformal superalgebras are also crucial for the construction of superconformal super Yang-Mills theories \cite{Brink:1976bc}. In \cite{Ferrara:1977ij}, it was pointed out that technical problems inherent to higher derivative actions appear when models that realize local scale, chiral, proper conformal, supersymmetry and internal $SU(N)$ transformations are constructed.

In this paper we introduce a generalized dual operator that allows us to construct geometric actions that correspond to a generalization of Yang-Mills theories for a dual operator that breaks certain symmetries of the (super-)algebra. Such operator can be found naturally in the superconformal algebra. These algebras have been used to construct related theories of matter \cite{Alvarez:2020qmy} and extended MacDowell-Mansouri supergravities \cite{Alvarez:2021qbu}.

Moreover, super Yang-Mills theories have received renewed attention recently due to their applications when computing scattering amplitudes \cite{Arkani-Hamed:2012zlh,Elvang:2013cua}. It is therefore interesting to formulate theories in terms of a generalized dual operator so that it becomes possible to transfer some of the tools that are used in the computation of scattering amplitudes of super Yang-Mills theories \cite{vanNieuwenhuizen:2004rh}.

Grand unified theories, with or without supersymmetry, based on the $SU(N)$ group, require $N>5$. The simplest example is the Georgy-Glashow $SU(5)$ model \cite{Georgi:1974sy,Langacker:1980js}. An $SU(5)$ model with softly broken supersymmetry was also proposed in \cite{Dimopoulos:1981zb} and, in order to solve the proton decay and the doublet-triplet splitting problems of the Georgy-Glashow $SU(5)$ model, the flipped $SU(5)$ model was proposed in \cite{Barr:1981qv}. Some years later, it was introduced the supersymmetric flipped $SU(5)$ model that produce hierarchical neutrino masses \cite{Antoniadis:1987dx,Ellis:1992zr,Ellis:1993ks,Ellis:1992nq}.

Attempts of embedding supersymmetric $SU(N)$ GUT models in the conformal superalgebra have to deal with the technical difficulties encountered in \cite{Ferrara:1977ij}. In this paper we define a dual operator embedded in $SU(2,2|N)$ that allows us to construct unified theories of gravity, Yang-Mills and matter. We study the field equations, integrability conditions and symmetry conditions. We also provide an explicit definition of the dual operator and we introduce the notion of self-dual forms with respect to the dual operator. In section \ref{explicitmodel} we provide the explicit form of the action. In section \ref{conclu} we summarize our results.

\section{The model}\label{model}

We formulate the model purely in terms of a $SU(2,2|N)$ connection. Following the procedure of \cite{Alvarez:2013tga,Alvarez:2020qmy, Alvarez:2021qbu}, we gauge and break some of the $SU(2,2|N)$ symmetries explicitly in the action. The generators of the ordinary conformal group, $\JJ_{a}, \KK_{a}, \JJ_{ab}$ and $\DD$, together with internal generators $\TT_I$ and $\ZZ$ of $SU(N)$ and $U(1)$ respectively, form the bosonic algebra. We have additionally $N$ complex spinorial supercharges, $\QQ_i^\alpha$ and $\QQb^i_\alpha$ (for the representation see appendix \ref{representationApp}). It is convenient for us to work with such Dirac supercharges because they make the full $R$-symmetry manifest \cite{Trigiante:2016mnt}, when the $R$-symmetry has been identified with the internal $SU(N)\times U(1)$. The gauge connection is given by
\begin{equation}\label{gaugeconnection}
\AAA= \Omega+\QQb_\alpha^i \se \psi_i^\alpha+\overline{\psi}_\alpha^i \se \QQ_i^\alpha\ \,,
\end{equation}
where 
\begin{equation}
\Omega= \half \omega^{ab}\JJ_{ab}+f^a \JJ_a+g^a\KK_a+h\DD+A^I \TT_I+A\ZZ \,.
\end{equation}
Curvatures are defined as usual,
\begin{equation}
 \FF = d\AAA +\AAA \wedge \AAA\,,
\end{equation}
where
\begin{equation}
\FF=\half \calF^{ab} \JJ_{ab}+\calF^a \JJ_a +\calG^a \KK_a +\calH \DD +\calF^I \TT_I +\calF \ZZ +\QQb \calX +\ocalX \QQ\,. 
\end{equation}
and the explicit form of the curvature components are given in appendix \ref{curvaturesApp}.

The action can be formulated as a generalization of the MacDowell-Mansouri action \cite{MacDowell:1977jt}
\begin{equation}\label{action}
 \mathcal{S}= - \int \langle \FF \circledast \FF \rangle\,.
\end{equation}
It is sufficient to have a dual operator $\circledast$ with the following requirements:
\begin{enumerate}[i.]
 \item it is demanded to have similar properties to the usual Hodge dual operator: linearity, parity odd and that $\circledast$ maps 2-forms valued on the algebra on to 2-forms valued on the algebra or in the complexified algebra at most. \label{prop1}
 \item That the operator $\circledast$ introduces a grading among the bosonic spacetime generators. We will denote the operator behind this grading by $S$, which is independent of the grading $\Gamma$, that defines fermionic and bosonic operators. The dual operator must satisfy $\circledast^2 = -1$ when projected to the corresponding subspace of the superalgebra. Below, in (\ref{Soperator}), we will make use of the latter property in order to uniquely define $S$. \label{prop-1}
\end{enumerate}
These properties will allow us to define actions with a gauge symmetry $G^+ \subset SU(2,2)\times SU(N) \times U(1)$, where $G^+$ is the group generated by the $S$-grading even generators. The symmetries generated by the $S$-grading odd generators, and the supersymmetries are going to be on-shell symmetries. These on-shell symmetries were discussed in \cite{Alvarez:2021qbu}, where for a supergravity model with $G^+ \subset SU(2,2)\times SU(2) \times U(1)$, where we pointed out that it is sufficient to demand a condition that is, in principle, weaker than the on-shell condition. See section \ref{symmetries}.

Additionally, \emph{a geometric coupling with matter can be implemented if the $S$-grading odd generators are dual to a set of fields that can be identified with a set of orthonormal frames}. By geometric we mean that the spin-1/2 fields carrying a representation of the $G$ group are going to be included in the gauge potential (\ref{gaugeconnection}). The resulting action (\ref{action}) contains the minimal coupling plus some extra gravitational nonminimal couplings that are tipically suppressed by powers of $M_P^{-2}$.

A dual operator with properties \ref{prop1} and \ref{prop-1} can be implemented in the conformal superalgebra $su(2,2|N)$. In order to do so we will use the spinor representation of the conformal group, where $\DD \sim \gf $, $\JJ_a \sim \gamma_a$, $\KK_a \sim \epsilon_a{}^{bcd}\gamma_{bcd}$ and $\JJ_{ab}\sim \Sigma_{ab}$, see appendix \ref{representationApp}. We will define the operator $S\sim \gf $ that introduces the following grading among bosonic generators 
\begin{equation}\label{Sgrading}
 \begin{tabular}{c|c}
  Even generators & Odd generators \\ \hline
  $[\JJ_{ab},S]=0$ & $\{\JJ_a,S\}=0$ \\
  $[\ZZ,S]=0$ & $\{\KK_a,S\}=0$ \\
  $[\TT_I,S]=0$ & \\
  $[\DD,S]=0$
 \end{tabular}
\end{equation}
where $\ZZ$ is the central generator of the superconformal group and $\TT_I$ are the generators of $SU(N)$. Moreover, the use of the spinor representation and the existence of $S$-grading odd generators is crucial to include matter fields in a geometric way. This is so because the fields dual to $\JJ_a$ and $\KK_a$ generators can be identified with the orthonormal frames,
\begin{equation}\label{auxiliaryfields}
 f^a = \rho e^a\,, \qquad g^a = \sigma e^a\,,
\end{equation}
where $\rho$ and $\sigma$ are integration constants. The identificaction is possible because the $S$-grading odd fields play the role of auxiliary fields in the action which can be fixed by solving their field equations and inserted back in the action. As shown in section 3.2, the identification \eqref{auxiliaryfields} corresponds to AdS or dS (or even Minkowski) vacua.  

After the fixing of the auxiliary fields, the resulting kinetic term for fermionic matter are Dirac (for $\rho$, $\sigma$ arbitrary), or chiral (when $\sigma=\pm \rho$, chiral in the sense that only one chirality is present). The explicit calculation is shown in section \ref{matter}.

Before defining the dual operator explictly, let us remark that the action (\ref{action}) can be thought of a Yang-Mills theory for an embedding $G^+ \hookrightarrow SU(2,2|N)$, where the horizontality condition is satisfied by the $S$-grading even generators only \cite{Trigiante:2016mnt}, denoted by $G^+$. To this end is crucial the existence of a dual operator with the following properties of the supertrace
\begin{align}
 &\langle \EE \circledast \OO \rangle = 0 = \langle \OO \circledast \EE \rangle\,,\label{EO}\\
 &\langle \EE_1 \circledast \EE_2 \rangle=\langle \EE_2 \circledast \EE_1 \rangle\,,\\
 &\langle \OO_1 \circledast \OO_2 \rangle=-\langle \OO_2 \circledast \OO_1 \rangle\,.\label{OO}
\end{align}
where $\OO$ represents two forms valued on $\circledast$-grading odd generators and $\EE$ represents two forms valued on $\circledast$-grading even generators, the latter being the $S$-grading even bosonic generators plus the supercharges, see table \ref{tablegradding}.

\begin{table}[h]
\begin{center}
\begin{tabular}{cl}
 gradding & domain    \\ \hline
 $\gf$    & $\underline{\JJ_a}, \underline{\KK_a} , \JJ_{ab}, \DD$    \\
 \textdownarrow & \\
 $S$    & $\underline{\JJ_a}, \underline{\KK_a} , \JJ_{ab}, \DD, \ZZ, \TT_I$     \\
  \textdownarrow & \\
$\circledast$ & $\underline{\JJ_a}, \underline{\KK_a} , \JJ_{ab}, \DD, \ZZ, \TT_I, \QQ, \QQb$   \\ 
\end{tabular}
\caption{Underlined are the odd generators w.r.t. each gradding operator. The spin 1/2 representation of the conformal algebra ($su(2,2)$) has a natural gradding defined by $\gf$. This embedding allows us to construct the grading $S$ of $su(2,2) \times su(N)$. With $S$ we can define the gradded dual operator $\circledast$ that acts on $su(2,2|N)$, that introduces the grading (\ref{EO})-(\ref{OO})  }\label{tablegradding}
\end{center}
\end{table}

The reason behind the cancellation of the kinetic term for $f^a$ and $g^a$ can be attributed to (\ref{OO}), that implies
\begin{equation}\label{proptrace1}
 \langle \FF^- \circledast \FF^- \rangle = 0\,,
\end{equation}
where $\FF^-$ are components of the curvature along the $S$-grading odd generators,
\begin{equation}
\FF^-=\calF^a \JJ_a +\calG^a \KK_a\,. 
\end{equation}
An important consequence of (\ref{EO}) is the fact that
\begin{equation}\label{proptrace2}
 \langle (\FF-\FF^-) \circledast \FF^- \rangle = 0=\langle \FF^- \circledast (\FF-\FF^-) \rangle\,,
\end{equation}
where $(\FF-\FF^-)$ contains all the components of the curvature except the $S$-grading odd generators. From (\ref{OO}), we get
\begin{equation}\label{proptrace3}
 \langle \FF_1^- \circledast \FF_2^- \rangle+\langle \FF_2^- \circledast \FF_1^- \rangle=0\,.
\end{equation}
In fact (\ref{proptrace3}) can be seen as a consequence of (\ref{proptrace1}) when $\FF_1^-=\FF^-$ and $\FF_2^-=\delta \FF^-$.

Using (\ref{proptrace1}), (\ref{proptrace2}) and (\ref{proptrace3}), it is easy to compute the change of the lagrangian (\ref{action}) under a general variation of the gauge connection,
\begin{equation}\label{genvar}
 \delta (-\langle \FF \circledast \FF \rangle )=-2 d \langle \delta \AAA \circledast (\FF-\FF^-)\rangle -2\langle \delta \AAA D_\AAA\circledast (\FF-\FF^-)\rangle\,.
\end{equation}

The first term on the r.h.s. of (\ref{genvar}) is a boundary term, that in fact provides the appropriate counter terms for defining charges in gravity. The second term provides the field equations of the model, which provided the fact that the supertrace is nondegenerate (see Appendix \ref{representationApp}, eqs. (\ref{trace1})-(\ref{trace-1})), can be stated simply as
\begin{equation}\label{fieldeqs}
 D_\AAA\circledast (\FF-\FF^-)=0\,.
\end{equation}
From (\ref{fieldeqs}) we can see that the fields in $\Omega^-$ are auxiliary fields. The fact that (\ref{auxiliaryfields}) is a consistent choice can be worked out by inspecting the explicit form of (\ref{fieldeqs}). Integrability conditions come from applying a covariant derivative to the three-form (\ref{fieldeqs}), where $D_\AAA \omega^{(3)}\equiv d\omega^{(3)}+\{ \AAA,\omega^{(3)}\}$ and $D^2_\AAA \omega^{(2)} = [\FF,\omega^{(2)}]$. Therefore the integrability condition is given by
\begin{equation}\label{intcond}
 [\FF,\circledast (\FF-\FF^-)]=0\,.
\end{equation}
These conditions determine relations between $\circledast$-odd curvatures and the fermionic curvatures that are satisfied on-shell. Relations (\ref{intcond}) do not impose restrictions on the $S$-grading even curvatures. As it is usual in Yang-Mills, gravity and supergravity theories, the integrability conditions are intimately related to the gauge invariance and therefore, in section \ref{symmetries}, we will provide the explicit form of the integrability conditions.

\subsection{Symmetries of the model}\label{symmetries}

The grading $S$ implies the splitting in unbroken and broken symmetries. Let us write the bosonic gauge connection as
\begin{equation}\label{omega+-}
 \Omega=\Omega^+ + \Omega^-\,,
\end{equation}
where
\begin{equation}
 \Omega^+= \half \omega^{ab}\JJ_{ab}+h\DD+A^I \TT_I+A\ZZ\,, \quad \Omega^-= f^a \JJ_a+g^a\KK_a\,.
\end{equation}
The gauge symmetries associated to the $S$-grading even generators are going to be genuine gauge symmetries, while the transformations associated to the $S$-grading odd generators are going to be conditional symmetries \cite{Alvarez:2021qbu}. Under a gauge transformation of the gauge potential, $\delta \AAA = D_\AAA G$, where
\begin{equation}
 G=G^++G^-+\QQb\epsilon -\overline{\epsilon}\QQ\,,
\end{equation}
the lagrangian changes by
\begin{equation}
 \delta (-\langle \FF \circledast \FF \rangle )=-2 d \langle D_\AAA G \circledast (\FF-\FF^-)+G D_\AAA \circledast (\FF-\FF^-)\rangle +2\langle G [\FF,\circledast (\FF-\FF^-)]\rangle\,.
\end{equation}
Thanks to the nondegenerate supertrace, the last term gives the same condition as the integrability conditions (\ref{intcond}), which can be seen as a generalization of an identity in Yang-Mills theories,
\begin{equation}\label{commYM}
[\FF, \ast \FF]=0\,, 
\end{equation}
to a model with certain number of broken symmetries. In a conventional Yang-Mills theory, the absence of $S$-grading odd generators means that (\ref{commYM}) is identically satisfied and therefore the whole symmetry group is gauged. In the present model the term $[\FF,\circledast (\FF-\FF^-)]$ expands along the $S$-grading odd generators and the supercharges only and therefore those symmetries are not gauge symmetries while the $G^+$ are going to be gauge symmetries. Note, however, that the condition $[\FF,\circledast (\FF-\FF^-)]$ coincides with the integrability condition of the field equations, and therefore $G^-\otimes (\QQb\epsilon -\overline{\epsilon}\QQ)$ are going to be on-shell symmetries.

When computing (\ref{intcond}) the following splitting of terms appear
\begin{align}
 [\FF,\circledast (\FF-\FF^-)]=&[\FF^+,\circledast\FF^+]+[\FF^-,\circledast \FF^+]+[\XX,\circledast\XX]\nonumber\\
 &+[(\FF-\XX),\circledast \XX]+[\XX,\circledast \FF^+]\,.\label{intcondv2}
\end{align}
where $\XX$ are the fermionic components of the curvature and $\FF^+=(\FF-\FF^--\XX)$ is a linear combination of curvatures along the $S$-grading even bosonic generators. Let us comment on the possible terms that appear in (\ref{intcondv2}). Terms of the form
\begin{equation}\label{intcondbos}
 [\FF^+,\circledast \FF^+]\equiv 0\,
\end{equation}
can be checked to vanish identically thanks to the properties of the $\circledast$ operator along $\FF^+$, as it happens to the usual Yang-Mills term and the gravity term in the MacDowell-Mansouri action.

Terms of the form
\begin{equation}\label{intcondXX}
[\XX,\circledast \XX] 
\end{equation}
pose some danger to the $G^+$ symmetries in the sense that they may contain $S$-grading even generators if the action of the operator $\circledast$ acting on fermionic components is not chosen appropriately. Using the embedding $S \sim \gf $ of $S$ in the superconformal algebra and the fact that
\begin{equation}\label{B-}
 [\{\QQ,\QQb \},\gf ] \sim B^-\,,
\end{equation}
where $B^-$ represents a linear combination of the $S$-grading odd generators, we can see that the definition
\begin{equation}\label{fermiondualchoice}
 \circledast (\QQb \calX) \sim \pm (\QQb i\gf \calX )\,,
\end{equation}
prevents the appearance of $\FF^+$ terms on the r.h.s. of (\ref{intcondXX}). Thus the choice (\ref{fermiondualchoice}), which is based on properties \ref{prop1} and \ref{prop-1}, implies
\begin{equation}\label{intcondXX2}
[\XX,\circledast \XX] \sim B^-\,.
\end{equation}
From (\ref{intcondbos}) and the absence of $B^+$ terms in (\ref{intcondXX2}), we can see that symmetries generated by $G^+$ are gauge symmetries. From (\ref{B-}) we also see that terms of the form $[\XX,\circledast \XX]$ contribute to the conditions on the $G^-$ transformations. Conditions on the $G^-$ transformations also come from $[\FF^-,\circledast \FF^+]$ terms.

Terms of the form
\begin{equation}\label{QQbarterms}
 [(\FF-\XX),\circledast \XX]+[\XX,\circledast \FF^+]\,,
\end{equation}
impose conditions along the susy transformations, which are generated by $\QQb\epsilon -\overline{\epsilon}\QQ$. In the next section we will define the operator $\circledast$ explicitly and derive the explicit form of the conditions (\ref{intcondv2}).

\subsection{Explicit form of the dual operator}

It is in our interest to keep the internal symmetry group, generated by $\TT_I$ and $\ZZ$, unbroken. This means that the right choice for the dual operator is the usual Hodge dual when acting on the $\calF^I$ components of the curvature 2-form, where
\begin{align}
\circledast \left( \calF^I \TT_I \right)=&\ast \left( \calF^I \TT_I \right)\nonumber\\
=&\ast  \calF^I \TT_I\,. 
\end{align}

In order to determine the form of the dual operator acting on the spacetime components of the curvature we need to introduce the $S$-grading. For the sake of definiteness let us define in the meantime the $S$-grading operator by
\begin{equation}\label{Soperator}
 S=i(\gf )^A_{\phantom{A}B}\,,
\end{equation}
where $A,B$ are indices in the superalgebra representation, see appendix \ref{representationApp}. It can be checked that $S^2=-(1)^A_{\phantom{A}B}$, when $A,B \in \alpha,\beta$ are projected in the spacetime block of the superalgebra, and that $S$ defines a grading (\ref{Sgrading}) among spacetime generators. Therefore a dual operator acting as $S$ in the spacetime block will imply properties (\ref{proptrace1})-(\ref{proptrace3}) on the bilinear invariants. Therefore, an appropriate choice for the action of $\circledast$ along $\half \calF^{ab} \JJ_{ab}+\calF^a \JJ_a +\calG^a \KK_a$ is
\begin{align}
 \circledast \left( \half \calF^{ab} \JJ_{ab}+\calF^a \JJ_a +\calG^a \KK_a\right)=&S \left( \half \calF^{ab} \JJ_{ab}+\calF^a \JJ_a +\calG^a \KK_a\right)\nonumber\\
 =&\half \calF^{ab} S\JJ_{ab}+\calF^a S\JJ_a +\calG^a S\KK_a\,.
\end{align}

The action of the dual operator on the components $\calH \DD +\calF \ZZ$ can be determined by demanding properties \ref{prop1} and \ref{prop-1}. From property \ref{prop1} we can discard the option $\circledast \sim S$ and therefore,
\begin{align}
\circledast \left( \calH \DD +\calF \ZZ \right)=&\ast \left( \calH \DD +\calF \ZZ \right)\nonumber\\
=&\ast \calH \DD +\ast \calF \ZZ\,. 
\end{align}

Finally, the action of the dual operator on the fermionic components $\QQb \calX +\ocalX \QQ$ can be defined as either $\sim i\gf $ or $\sim \ast$ since both operators satisfy properties \ref{prop1} and \ref{prop-1},
\begin{align}
\circledast \left( \QQb \calX +\ocalX \QQ\right)=&\ast \left( \QQb \calX +\ocalX \QQ \right)\nonumber\\
=& \QQb \ast\calX +\ast\ocalX \QQ\,. \label{circledastfermion1}
\end{align}
or
\begin{equation}\label{circledastfermion2}
\circledast \left( \QQb \calX +\ocalX \QQ\right)= i\QQb \gf \calX +i\ocalX \gf \QQ\,. 
\end{equation}
The fact that the action of $S$ and $\ast$ is the same when actiong on $e^a e^b \Sigma_{ab}$ is a crucial property in order to construct the fermionic kinetic term. One advantage of using the representation $i\gf $ of the dual operator is that it makes the separation of unbroken and broken symmetries manifest by implying properties (\ref{EO}) and (\ref{OO}), also see next section. Another reason for choosing (\ref{circledastfermion2}) is of dynamical character, see discussion around (\ref{tracespacetime}).

There is only one ambiguity left by properties \ref{prop1} and \ref{prop-1}, which is the sign choice of the action of $\circledast$ along the different components of the curvature. Therefore we will define the dual operator as
\begin{align}
\circledast \FF=&(\varepsilon_s S)\left(\half \calF^{ab} \JJ_{ab}+\calF^a \JJ_a +\calG^a \KK_a\right)\nonumber\\
&+(\varepsilon_1\ast)\calH \DD +(\varepsilon_2\ast) \calF^I \TT_I +(\varepsilon_3\ast)\calF \ZZ + \QQb (-i\varepsilon_\psi\gf ) \calX + \ocalX(-i\varepsilon_\psi\gf ) \QQ\,.\label{dualop}
\end{align}
In (\ref{dualop}) $\varepsilon_s$, $\varepsilon_1$, $\varepsilon_2$, $\varepsilon_3$ and $\varepsilon_\psi$ take the values +1 or -1. The ambiguity in $\varepsilon_s$, $\varepsilon_1$, $\varepsilon_2$ and $\varepsilon_3$ can be removed by demanding the correct sign of the kinetic terms of the bosonic fields in the action, however the fermionic kinetic term do not have definite sign and therefore the ambiguity in $\varepsilon_\psi$ cannot be removed by such requirement. The choice of sign in $\varepsilon_1$, $\varepsilon_2$ and $\varepsilon_3$ will depend on the details of the superalgebra representation, and in the more subtle case of $\varepsilon_s$, the sign choice will also depends on the sector of the gravity theory that is going to be used as vaccuum, see section \ref{gravity}. The choice of sign in $\varepsilon_\psi$ though will imply the exact cancellation or not of Pauli-like couplings with the bosonic curvatures $R^{ab}$, $H$, $F^I$ and $F$. A Pauli-like coupling with the curvature $R^{ab}$ induce a direct coupling of matter with the Riemann tensor.

Let us discuss now certain dynamical aspects. With (\ref{Soperator}), we can see that the term
\begin{equation}\label{tracespacetime}
 \langle \left(\half \calF^{ab} \JJ_{ab}+\calF^a \JJ_a +\calG^a \KK_a \right) S \left(\half \calF^{ab} \JJ_{ab}+\calF^a \JJ_a +\calG^a \KK_a \right) \rangle \,,
\end{equation}
where $\langle \ {}_{\cdot} \ \rangle$ stands for a supertrace. The choice (\ref{Soperator}) in (\ref{tracespacetime}) cancels out the kinetic terms of fields $f^a$ and $g^a$ and therefore these fields become auxiliary fields. A posteriory inspection of field equations shows that the fixing (\ref{auxiliaryfields}) is in fact a consistent choice. In the gravity sector, after fixing the auxiliary fields by (\ref{auxiliaryfields}) in (\ref{tracespacetime}), we recover Einstein-Hilbert, cosmological constant and Gauss-Bonnet terms.

It is noteworthy that the fermion kinetic term will depend on the action of $\circledast$ after auxiliary fields are fixed by (\ref{auxiliaryfields}), and therefore it dependes on the action of $\circledast$ on the subspace of two forms valued on the antisymmetric Clifford algebra $\gamma_{ab}$, so (\ref{circledastfermion1}) and (\ref{circledastfermion2}) give the same result,
\begin{equation}
 *(\se \se)=i\gf \se \se\,.
\end{equation}
However, extra terms with respect to the Dirac term and a different structure of nonminimal couplings could appear. This reason adds weight to the choice (\ref{circledastfermion2}).

With the definition of $\circledast$ we can carry out the explicit calculation of (\ref{intcondv2}) using (\ref{dualop})gives
\begin{align}
 [\FF,\circledast (\FF-\FF^-)]=&(\calG^a (\varepsilon_1\ast)\calH -\varepsilon_s\frac{1}{2}\epsilon^a{}_{bcd}\calF^b\calF^{cd}+ \ocalX\gamma^a(-i\varepsilon_\psi\gf )\calX)\JJ_a\nonumber\\
 &+(\calF^a (\varepsilon_1\ast)\calH -\varepsilon_s\frac{1}{2}\epsilon^a{}_{bcd}\calG^b\calF^{cd}-\ocalX\tgamma^a(-i\varepsilon_\psi\gf )\calX)\KK_a\nonumber\\
 &+[(\FF-\XX),\circledast \XX]+[\XX,\circledast \FF^+]\,.\label{intcondexp}
\end{align}
The last line is proportional to fermionic terms only (see (\ref{QQbarterms})),
\begin{equation}\label{square}
 [(\FF-\XX),\circledast \XX]+[\XX,\circledast \FF^+]=\QQb \left(  \calF^r  \rho(B_r)\circledast \calX-\circledast((\calF^+)^r  \rho(B^+_r))\calX \ \right) +\text{C.C.}\,,
\end{equation}
where C.C. stands for complex conjugated, $B_r$ represents all the bosonic generators, $\calF^r$ the curvature components along $B_r$ and $(B^+)_r$ the $S$-grading even generators. The notation $\rho(B_r)$ corresponds to the spin-1/2 representation if the generator $B_r$. The term in parenthesis in (\ref{square}) is given by,
\begin{align}
 \left(  \calF^r  \rho(B_r)\circledast \calX-\circledast((\calF^+)^r  \rho(B^+_r))\calX \ \right)=& \left[\left((-i\varepsilon_\psi\gf ) -(i\varepsilon_s\gf )\right)\left(\half\calF^{ab}\Sigma_{ab}\right)\right]\calX \nonumber\\
 &+ \left[\left( (-i\varepsilon_\psi\gf ) -(\varepsilon_1\ast)\right) \left(\half\calH\gf \right)\right]\calX\nonumber\\
 &+ \left[\left( (-i\varepsilon_\psi\gf ) -(\varepsilon_2\ast)\right)\left(-\frac{i}{2} \calF^I\lambda_I\right)\right]\calX\nonumber\\
 &+\left[ \left( (-i\varepsilon_\psi\gf ) -(\varepsilon_3\ast)\right)\left(-iz(4/N-1)\calF\right)\right]\calX\nonumber\\
 &+\left( \half \calF^a\gamma_a +\half \calG^a\tgamma_a  \right)(-i\varepsilon_\psi\gf )\calX\,.\label{intcondsusy}
\end{align}
No conditions appear along the $S$-grading even generators thanks to the fact that (\ref{intcondexp}) expands along $\JJ_a$, $\KK_a$, $\QQb$ and $\QQ$ only. As mentioned above, the meaning of this is that $G^+$ are gauge symmetries of the model. However, in (\ref{intcondexp}) individual components of the $\FF^+$ curvatures do appear multiplying components of $\FF^-$ and $\XX$ curvatures to form quadratic combinations. Therefore we can see that $G^-$ transformations and supersymmetric transformations can be symmetries if the bosonic background vacuum is flat
\begin{equation}\label{bosonicF+}
 \FF^+ = 0\,, \quad \text{when} \ \Psi=0\,.
\end{equation}
Besides (\ref{bosonicF+}), there are a few noteworthy facts that stem from (\ref{intcondexp}). Firstly, there may exist bosonic backgrounds such that $\FF^-=0$ with or without vanishing $\FF^+$ that are $G^-$ and susy invariant. Secondly, there may exist  backgrounds with vanishing $\FF^+$ but non vanishing $\FF^-$ such that the last line of (\ref{intcondsusy}) vanishes, that condition can be satisfied if $(-i\varepsilon_\psi\gf )\calX$ lives in the kernel of the linear operator $\left( \half \calF^a\gamma_a +\half \calG^a\tgamma_a  \right)$. Thirdly, $\varepsilon_\psi$ can be chosen such that one or more lines of the 1st to 4st lines of (\ref{intcondsusy}) identically vanish, and therefore opening the possibility of realizing susy on non-flat $\FF^+$ background curvatures.

It is noteworthy that conditions (\ref{intcondexp}) can be obtained also after a lengthy computation based on the symmetry transformations acting on the fields (given in appendix \ref{symApp}) and the explicit form of the action (\ref{action}). However the usage of the properties of the dual operator greatly simplifies the computation.

\subsection{Self-dual two-forms}

Lastly, the definition of the dual operator provides us with a concrete notion of self dual or anti-self dual solutions,
\begin{equation}\label{selfdual}
 \circledast (\FF-\FF^-) = \pm (\FF-\FF^-)\,.
\end{equation}
Such self-duality condition is motivated on the definition of the operator $\circledast$ and the form of the field equations, see (\ref{fieldeqs}). We can see that such self dual curvature is a solution of the field equations. Condition (\ref{selfdual}) along the components of the curvature gives
\begin{align}
 &\half\varepsilon_s \epsilon^{ab}{}_{cd}\calF^{cd}=\pm\half\calF^{ab}\,,\\
 &\varepsilon_1\ast \calH=\pm \calH\,,\\
 &\varepsilon_2\ast \calF^I=\pm \calF^I\,,\\
 &\varepsilon_3\ast \calF=\pm \calF\,,\\
 &(-i\varepsilon_\psi \gf) \calX=\pm\calX\,,\label{selfdualcalX}\\
 &\ocalX(-i\varepsilon_\psi \gf) =\pm\ocalX\,,\label{selfdualocalX}
\end{align}
Conditions (\ref{selfdualcalX}) and (\ref{selfdualocalX}) can be satisfied for fermions with well defined chiral handedness. In a purely bosonic background the self dual and anti-self dual conditions reduce to
\begin{align}
 &\half\varepsilon_s \epsilon^{ab}{}_{cd}\calR^{cd}=\pm\half\calR^{ab}\,,\\
 &\varepsilon_1\ast (H+f^ag_a)=\pm (H+f^ag_a)\,,\\
 &\varepsilon_2\ast F^I=\pm F^I\,,\\
 &\varepsilon_3\ast F=\pm F\,.
\end{align}

\section{Explicit form of the action}\label{explicitmodel}

Using the properties of the supertrace (\ref{proptrace1})-(\ref{proptrace3}) we can expand the lagrangian as
\begin{equation}\label{lag}
-\langle \FF \circledast \FF\rangle=-\langle \FF^+ \circledast \FF^+\rangle-\langle \XX \circledast \XX\rangle\,.
\end{equation}

In this section we will describe the physical content of (\ref{lag}).

\subsection{Field equations}

Field equations can be obtained by direct variation of the Lagrangian (\ref{actionexpand}) with respect to the fundamental fields $\delta \omega^{ab}$, $\delta h$, $\delta A^I$, $\delta A$, $\delta f^a$, $\delta g^a$, $\delta (\se \psi)$ and , $\delta (\psibar \se)$ or, equivalently, by using expression (\ref{fieldeqs}),
\begin{align}
 0=D\circledast (\FF-\FF^-)\equiv & d\circledast(\FF-\FF^-)+[\AAA,\circledast(\FF-\FF^-)]\,\nonumber\\
 =&D_\Omega\circledast \FF^+ + D_\Omega \circledast \XX +[\Psi,\circledast \FF^+]+[\Psi,\circledast \XX]\,,
\end{align}
where $\Psi=\QQb\se\psi+\psibar\se \QQ$. Along the generators $\JJ_{ab}$, $\DD$, $\TT_I$, $\ZZ$, $\JJ_a$, $\KK_a$ and  $\QQb$, we obtain,
\begin{align}
 0=&\half \varepsilon_s \epsilon^{ab}{}_{cd}D_\omega \calF^{cd} -\ocalX (-i\varepsilon_\psi \gf)\left(-\half \Sigma^{ab}\right)\se \psi +\psibar \se \left(-\half \Sigma^{ab}\right)(-i\varepsilon_\psi \gf)\calX\,,\label{fieldeqJab}\\
 0=&d(\varepsilon_1\ast \calH)-\ocalX (-i\varepsilon_\psi \gf)\left(\half \gf\right)\se \psi +\psibar \se \left(\half \gf\right)(-i\varepsilon_\psi \gf)\calX\,,\\
 0=& D_{(A^J \TT_J)}(\varepsilon_2\ast \calF^I) -\ocalX (-i\varepsilon_\psi \gf)\left(-i\lambda_I\right)\se \psi +\psibar \se \left(-i\lambda_I\right)(-i\varepsilon_\psi \gf)\calX\,,\\
 0=&d(\varepsilon_3\ast \calF)-\ocalX (-i\varepsilon_\psi \gf)\left(-\frac{i}{4z}\right)\se \psi +\psibar \se \left(-\frac{i}{4z}\right)(-i\varepsilon_\psi \gf)\calX\,,\\
 0=&g^a(\varepsilon_1\ast \calH)-\half\varepsilon_s\epsilon^a{}_{bcd}f^b \calF^{cd}-\ocalX (-i\varepsilon_\psi \gf)\left(\half\gamma^a\right)\se \psi +\psibar \se \left(\half\gamma^a\right)(-i\varepsilon_\psi \gf)\calX\,,\\
 0=&f^a(\varepsilon_1\ast \calH)-\half\varepsilon_s\epsilon^a{}_{bcd}g^b \calF^{cd}-\ocalX (-i\varepsilon_\psi \gf)\left(-\half\tgamma^a\right)\se \psi +\psibar \se \left(-\half\tgamma^a\right)(-i\varepsilon_\psi \gf)\calX\,,\\
 0=&D_\Omega (-i\varepsilon_\psi \gf)\calX-\left(\half\calF^{ab}(i\varepsilon_s \gf)\Sigma_{ab}+\half\gf(\varepsilon_1\ast \calH)\right.\nonumber\\
 &\qquad\qquad\qquad\qquad\qquad\left.-\frac{i}{2}\lambda_I(\varepsilon_2\ast \calF^I)-iz\left(\frac{4}{N}-1\right)(\varepsilon_3\ast \calF)\right)\se\psi\,,\label{fieldeqQQb}
\end{align}
respectively. In (\ref{fieldeqJab}) we used $SJ_{ab}=(1/2)\epsilon_{ab}{}^{cd}J_{cd}$. The field equation along the $\QQ$ generator is the complex conjugate of (\ref{fieldeqQQb}).

In a purely bosonic background the field equations take the simplified form
\begin{align}
 0= &D_\Omega\circledast \FF^+ \,,
\end{align}
Along the $\JJ_{ab}$, $\DD$, $\TT_I$, $\ZZ$, $\JJ_a$ and $\KK_a$ generators we obtain,
\begin{align}
 0=&\half \varepsilon_s \epsilon^{ab}{}_{cd}D_\omega \calR^{cd}\,,\\
 0=&d(\varepsilon_1\ast (H+ f^b g_b))\,,\\
 0=& D_{(A^J \TT_J)}(\varepsilon_2\ast F^I)\,,\\
 0=&d(\varepsilon_3\ast F)\,,\\
 0=&g^a(\varepsilon_1\ast (H+ f^b g_b))-\half\varepsilon_s\epsilon^a{}_{bcd}f^b \calR^{cd}\,,\\
 0=&f^a(\varepsilon_1\ast (H+ f^b g_b))-\half\varepsilon_s\epsilon^a{}_{bcd}g^b \calR^{cd}\,.
\end{align}
The field equations are solved in flat backgrounds and self dual backgrounds when the self duality condition (\ref{selfdual}) can be satisfied. As an illustration and as a viability test towards the construction of quasi-realistic models, in the next section we will make contact with General Relativity.

\subsection{Gravity terms and background vacuum}\label{gravity}

Working with the representation given in appendix \ref{representationApp} the lagrangian (\ref{lag}) can be written as
\begin{equation}\label{actionexpand}
-\langle \FF \circledast \FF\rangle=\frac{1}{4}\varepsilon_s\epsilon_{abcd}\calF^{ab}\calF^{cd}-\varepsilon_1 \calH\ast\calH-\half\varepsilon_2 \calF^I\ast \calF^I -4z^2\left(\frac{4}{N}-1\right) \varepsilon_3 \calF\ast \calF-2i\varepsilon_\psi \ocalX \gf \calX\,.
\end{equation}

The bosonic sector of the theory is given by setting $\psi=0$ on (\ref{actionexpand}),
\begin{equation}\label{actionexpandbos}
\calL_\text{bos}=\frac{1}{4}\varepsilon_s\epsilon_{abcd}\calR^{ab}\calR^{cd}-\varepsilon_1 (H+f^a g_a)\ast (H+f^b g_b)-\half\varepsilon_2 F^I\ast F^I -4z^2\left(\frac{4}{N}-1\right) \varepsilon_3 F\ast F\,.
\end{equation}
We can fix $\varepsilon_1=+1=\varepsilon_2$, but keeping them helps in tracking down the origin of terms. The value of $\varepsilon_3$, however, depends on $N$
\begin{equation}
 \varepsilon_3=\begin{cases}
  +1 \quad \text{for} \quad N<4\\
  -1 \quad \text{for} \quad N>4
 \end{cases}\,.
\end{equation}

Variations of (\ref{actionexpandbos}) with respect to $f^a$ and $g^a$ give us
\begin{align}
 \frac{\delta \calL_\text{bos}}{\delta f^a}=\varepsilon_s \epsilon_{abcd}f^b\calR^{cd}-2\varepsilon_1 g_a\ast(H+f^b g_b)\,,\\
 \frac{\delta \calL_\text{bos}}{\delta g^a}=-\varepsilon_s \epsilon_{abcd}g^b\calR^{cd}+2\varepsilon_1 f_a\ast(H+f^b g_b)\,,
\end{align}
from which we see that $f^a \sim g^a$, $H=0$ and $\calR^{ab}=0$ solves the equations. Using (\ref{auxiliaryfields}) with constant $\rho$ and $\sigma$ (but $\rho \ne \sigma$) in $\calR^{ab}$ we can see that AdS or dS vacua,
\begin{equation}
 R^{ab} \pm \ell^{-2}e^a e^b=0\,,
\end{equation}
are recovered, where,
\begin{equation}
 \ell^{-2}=|\rho^2-\sigma^2|\,, \quad \pm=\begin{cases}
                                             +1 \quad \text{for} \quad \rho >\sigma\\
                                             -1 \quad \text{for} \quad \rho <\sigma\\
                                          \end{cases}\,.
\end{equation}

The sign $\varepsilon_s$ and the scale of $\rho$ and $\sigma$ are fixed by comparing with the Einstein-Hilbert term and the cosmological constant term,
\begin{align}
 \frac{\varepsilon_s}{4}\int\epsilon_{abcd}\mathcal{R}^{ab}\mathcal{R}^{cd}=\frac{1}{16\pi G_N}\int d^4x |e| (R-2\Lambda)\,,
\end{align}
From here we read Newton constant and the cosmological constant
\begin{align}
 \frac{1}{16\pi G_N}=&\varepsilon_s (\rho^2-\sigma^2)\,,\\
 \Lambda=&-3(\rho^2-\sigma^2)\,.
\end{align}
If we take $\varepsilon_s=+1$, positivity of Newton constant implies $\rho >\sigma$ and therefore $\Lambda<0$, that is AdS vacuum. For $\varepsilon_s=-1$, the same argument implies $\rho <\sigma$ and therefore $\Lambda>0$, that is dS vacuum. The present model provides an example of a model with benign ghosts, since for a given value of $\varepsilon_s=+1\ (-1)$ there is always a ghost free sector of the theory based on the right choice for the auxiliary fields (see review \cite{Damour:2021fva}).

The sector in which $\rho$ and $\sigma$ are constant is the analogous of the ``unitary'' gauge of finite superconformal theories, that permits to recover Einstein gravity \cite{Fradkin:1985am}. It is interesting to consider a more general sector in which $\rho$ and $\sigma$ are scalar fields whose expectation value breaks spontaneously the conformal symmetry \cite{Englert:1976ep,Zee:1978wi}. Such sector contains a term in the lagrangian $|e| \rho^2 R \subset \calL$ typical of scalar-tensor theories in the Jordan frame, and this term induces a kinetic term for the scalar in the Einstein frame. Spontaneous breaking of Weyl invariance requires the absence of conformal canomalies, the study of which is beyond the scope of this paper but we consider an interesting study for the future.

\subsection{Matter action}\label{matter}

The fermion kinetic term is contained in the last term of (\ref{lag}), which comes from $-\langle \XX \circledast \XX \rangle$. Let us use the covariant derivative for the $\Omega$ connection to rewrite this term as follows,
\begin{align}
-\langle \XX \circledast \XX \rangle=&2\ocalX (-i\varepsilon_\psi\gf) \calX \nonumber\\
=&-2(\psibar \se) \lD_\Omega(-i\varepsilon_\psi\gf) D_\Omega(\se\psi)\,.\label{fermionterm}
\end{align}
As stated in section \ref{model}, the operator $\circledast=(-i\varepsilon_\psi\gf)$ satisfy properties \ref{prop1} and \ref{prop-1}, and the fields dual to the $S$-grading odd generators can be identified with a set of orthonormal frames. Such identificaction, provided in (\ref{auxiliaryfields}), is compatible with the structure of the gauge symmetries $G^+$.

The term (\ref{fermionterm}) contains quadratic derivatives of the spinor which is problematic from the dynamical point of view. However, the fact that $\circledast$ defines $S$-grading even generators implies that such quadratic derivatives form a boundary term. There remain first order derivatives of the spinor that provide the conventional kinetic term with minimal coupling of a spinor when (\ref{auxiliaryfields}). In order to appreciate these facts it is convenient to express (\ref{fermionterm}) using $D_\Omega=D^+ + \Omega^-$, where $D^+=d+\Omega^+$ and $\Omega^+$ is defined in (\ref{omega+-}). After rearranging terms we obtain
\begin{align}
2\ocalX (-i\varepsilon_\psi\gf) \calX=&-2i\varepsilon_\psi\left[\psibar(\lD^+ \se\gf \Omega^- \se-\se\Omega^- \gf\se D^+)\psi\right.\nonumber\\
&+\psibar(\sT\gf \Omega^- \se+ \se\Omega^- \gf\sT)\psi\nonumber\\
&\left.+\psibar \se\gf (D^+)^2\se\psi+\psibar\se \Omega^- \gf \Omega^- \se\psi \right]\nonumber\\
&-2i\varepsilon_\psi d[\psibar \se\gf D^+\se\psi]\,.\label{fermioniccurvatureterm}
\end{align}
The first line produces minimal coupling terms provided (\ref{auxiliaryfields}). Second and third lines are nonminimal coupling terms, which are provided in tensor notation in section \ref{modelsummary}.

In the first line of (\ref{fermioniccurvatureterm}) we can use
\begin{equation}\label{identityh}
 \psibar(\lD^+ \se\gf \Omega^- \se-\se\Omega^- \gf\se D^+)\psi=\psibar(\lD \se\gf \Omega^- \se-\se\Omega^- \gf\se D)\psi
\end{equation}
where $D=d+W$, where $W=\Omega^+-h\DD$, since the $h$ terms cancel out from such term thanks to the fact that $[\gf,\DD]=0$. The resulting covariant derivative for the spinor includes only $W$ terms
\begin{equation}\label{Wgaugeconnection}
W=\half \omega^{ab}\JJ_{ab}+A^I\TT_I+A\ZZ \,.
\end{equation}
The r.h.s. of (\ref{identityh}) is invariant under dilatation transformations thanks to the fact that $\Omega^-$ transforms non trivially under $G=\tau \DD$ transformations. Such symmetry though, is broken in a background vacuum with constant $\rho$ and $\sigma$. In a dilatation symmetry broken phase we can express (\ref{fermioniccurvatureterm}) in terms of the covariant derivative for the $W$ connection for simplicity.

Provided the auxiliary fields are given by (\ref{auxiliaryfields}), we get
\begin{equation}\label{omegaminusdirac}
 \Omega^-=\se \Pi(\alpha,\beta)\,, \quad \Pi(\alpha,\beta)=\alpha P_R+\beta P_L\,,
\end{equation}
where $P_{R,L}=\half (1\pm \gf)$ are right or left chiral projectors, and
\begin{equation}
 \alpha =\frac{\rho+\sigma}{2}\,, \quad \beta =\frac{\rho-\sigma}{2}\,.
\end{equation}

Using (\ref{omegaminusdirac}), the first line in (\ref{fermioniccurvatureterm}) is given by
\begin{equation}\label{Dirac}
 -2i\varepsilon_\psi \psibar(\lD \se\gf \Omega^- \se-\se\Omega^- \gf\se D)\psi=\frac{\varepsilon_\psi}{2}|e|d^4 x \psibar' (\overleftarrow{\slashed{\nabla}}-\slashed{\nabla})\psi'
\end{equation}
where $\psi'=\psi'_L+\psi'_R$ is a four component Dirac spinor with canonical physical units, and
\begin{equation}\label{psiscaling}
\psi'_L=\sqrt{24\alpha}P_L \psi\,, \quad \psi'_R=\sqrt{24\beta}P_R \psi \,.
\end{equation}
are four component chiral fermions (one for each $i$ index that is hidden). The Dirac adjoint spinor is given by $\psibar'=\psibar'_R +\psibar'_L$, where
\begin{equation}\label{psibarscaling}
\psibar'_L=\sqrt{24\alpha}\psibar' P_R\,, \quad \psibar'_R=\sqrt{24\beta}\psibar' P_L \,.
\end{equation}
Our conventions are consistent with $\psibar'_{R,L}\equiv\overline{\psi'_{R,L}}$. In (\ref{Dirac}) we have the covariant derivatives
\begin{align}
 (\nabla_\mu)^{\alpha j}_{i \beta}&=\delta^j_i \delta^\alpha_\beta \partial_\mu +\half \omega_\mu^{ab}\delta^j_i(\Sigma_{ab})^\alpha_{\ \beta}-\frac{i}{2}A_\mu^I(\lambda_I)_i^{\ j}\delta^\alpha_\beta-iz\left(\frac{4}{N}-1\right)A_\mu\delta^j_i \delta^\alpha_\beta\,,\label{covariantD}\\
 (\overleftarrow{\nabla}_\mu)^{\alpha j}_{i \beta}&=\overleftarrow{\partial}_\mu\delta^j_i \delta^\alpha_\beta -\half \omega_\mu^{ab}\delta^j_i(\Sigma_{ab})^\alpha_{\ \beta}+\frac{i}{2}A_\mu^I(\lambda_I)_i^{\ j}\delta^\alpha_\beta+iz\left(\frac{4}{N}-1\right)A_\mu\delta^j_i \delta^\alpha_\beta\,.
\end{align}


Before finishing this section let us give the form of certain nonminimal coupling terms of the spinor contained in (\ref{actionexpand}).
\begin{align}
 \calL_\text{NMC}(\psi^2)=&-2i \varepsilon_\psi \psibar(\sT\gf\Om\se+\se\Om\gf\sT)\psi\nonumber \\
 &+i(\varepsilon_\psi-\varepsilon_s)\psibar(f^af^b+g^ag^b)\gf \Sigma_{ab}\se\se\psi+i(\varepsilon_\psi+\varepsilon_1)f^a g_a\psibar \se\se\psi\nonumber\\ 
 &+2i\psibar \gf \se\left( \half(\varepsilon_\psi+\varepsilon_s)R^{ab}\Sigma_{ab}-\frac{i}{2}(\varepsilon_\psi-\varepsilon_2)F^I\lambda_I\right.\nonumber\\
 &\qquad \qquad \qquad \left.-iz(\varepsilon_\psi-\varepsilon_3)\left(\frac{4}{N}-1\right)F+\half (\varepsilon_\psi-\varepsilon_1)H\gf \right) \se \psi\,.\label{LNMCpsi2}
\end{align}
These terms come from the quadratic terms of the form $F^+_r\circledast (\psibar \rho(B^+_r)\psi)$ that are contained in the $\FF^+$ curvatures and from the second and third line of (\ref{fermioniccurvatureterm}). We kept the parameters $\varepsilon_x$ that help in tracking down the origin of each term.

In (\ref{actionexpand}) we also have quartic fermion terms. Such terms take the form $$ \text{const} \times \dV \ \phi^{UI} \phi_{UI} $$ where $\phi_{UI}=\psibar \gamma_U\lambda_I \psi$, where $\gamma_U$ is an element of Clifford algebra. These terms come from $(\psibar \rho(B^+_r)\psi)\circledast (\psibar \rho(B^+_r)\psi)$ that are contained in $\langle\FF^+\circledast \FF^+\rangle$. Collecting all such terms we obtain,
\begin{equation}
 \calL(\psi^4)/(\dV)= \half\left(\varepsilon_3\left(\frac{4}{N}-1\right) -\varepsilon_1\right)\phi^{ab}\phi_{ab}+6\varepsilon_s(\phi^2+\phi_5^2)+2\varepsilon_2 \phi^{abI}\phi_{abI}\,.
\end{equation}

\subsection{Summary of the model}\label{modelsummary}

In this section we provide a general procedure to obtain a phenomenological action in the sector (\ref{auxiliaryfields}) with constant $\rho$ and $\sigma$. It can be summarized by the following four steps: firstly, we set conditions (\ref{auxiliaryfields}) with constant $\rho$ and $\sigma$. Secondly, we pass to the physical gauge fields by canonically normalizing their kinetic terms. Thirdly, we define the physical spinors by implementing the scalings (\ref{psiscaling}) and (\ref{psiscaling}). Finally, we introduce a global rescale of the action with a phenomenological parameter.

The gauge couplings are defined once the Yang-Mills terms are canonically normalized,
\begin{equation}
 - a F\ast F = -\half F'\ast F'\,,
\end{equation}
implying that the physical gauge potential is $A'=\sqrt{2a}A$. Assuming that the covariant derivative acting on the spinor takes the form
\begin{equation}
 D = d-i g_0 \ \rho(T_r)A^r\,,
\end{equation}
where $g_0$ are the coupling constants implied by the superconformal algebra, see (\ref{covariantD}), we can deduce that with respect to the physical gauge potential $A'$, we have the coupling constant $g=g_0/\sqrt{2a}$. In the present model we obtain
\begin{align}
 g^{(SU(N))}=&1\,,\\
 g^{(U(1))}=&\text{sign}(4/N-1)\sqrt{\frac{\varepsilon_3}{8}\left(\frac{4}{N}-1\right)}\,,
\end{align}
where $\varepsilon_3$ is negative for $N>4$ and positive for $N<4$. The case $N=4$ is special because the superconformal algebra degenerates making the field $A\ZZ \subset \AAA$ irrelevant, i.e. $A$ completely disappears from the action for $N=4$.

Relations (\ref{psiscaling}) and (\ref{psiscaling}) scale each chiral components differently, implying that spinor bilinears that do not mix chiralities remain invariant and bilinears that mix chiralities scale according to the following rule
\begin{equation}\label{chiralscaling}
 \phi_U\equiv \psibar \gamma_U \psi=\begin{cases}
  \psibar'\gamma_U\psi' \quad \text{if} \quad \gamma_U=\gamma_a, \tgamma_a\\
  \frac{1}{24\sqrt{\alpha\beta}}\psibar'\gamma_U\psi' \quad \text{if} \quad \gamma_U=\mathds{1},\gamma_5,\gamma_{ab}\\
 \end{cases}
\end{equation}

Finally, a phenomenological parameter can be introduced by implementing a global scaling of the action
\begin{equation}\label{globalscaling1}
 \calL \rightarrow \calL'=\xi \calL\,.
\end{equation}
Such parameter allows us to set the scale of the cosmological constant or the gauge couplings. Under such scaling we need to rescale the gauge fields and the spinor, accordingto the rule
\begin{equation}\label{globalscaling2}
 A''=\sqrt{\xi}A'\,,\quad \psi''=\sqrt{\xi}\psi'\,,\quad \psibar''=\sqrt{\xi}\psibar'\,.
\end{equation}
As a result the gauge couplings and the cosmological constant get rescaled. Nonminimal couplings of the second line in (\ref{LNMCpsi2}) get rescaled and the rest do not get rescaled, see table \ref{tablecc}.

We present the action dropping primes for all fields, assuming that the action is written in terms of the physical fields. The final action is given by
\begin{eqnarray}\label{EYC}
 \cal{S} &=&  \int \dV \left[\frac{1}{16\pi \GN}(R-2\Lambda)-\frac{1}{4}H^{\mu\nu}H_{\mu\nu}-\frac{1}{4}F^{I\mu\nu}F^{I}_{\mu\nu}-\frac{1}{4}F^{\mu\nu}F_{\mu\nu} \right.   \nonumber\\
   && \left. +\frac{1}{2}
  \psibar (\lsD-\sD)\psi+T_\text{NMC}\right]
\end{eqnarray}
where $T_\text{NMC}$ stands for nonminimal coupling terms and $D$ is the covariant derivative for the $SO(3,1)\times SU(N)\times U(1)$ gauge connection,
\begin{align}
 (D)^{\alpha j}_{i \beta}&=\delta^j_i \delta^\alpha_\beta d +\half \omega^{ab}\delta^j_i(\Sigma_{ab})^\alpha_{\ \beta}-\frac{i}{2}g^{(SU(N))}A^I(\lambda_I)_i^{\ j}\delta^\alpha_\beta-i g^{(U(1))}A\delta^j_i \delta^\alpha_\beta\,,\label{covD}\\
 (\overleftarrow{D})^{\alpha j}_{i \beta}&=\overleftarrow{d}\delta^j_i \delta^\alpha_\beta -\half \omega^{ab}\delta^j_i(\Sigma_{ab})^\alpha_{\ \beta}+\frac{i}{2}g^{(SU(N))}A^I(\lambda_I)_i^{\ j}\delta^\alpha_\beta+ig^{(U(1))}A\delta^j_i \delta^\alpha_\beta\,.\label{covDl}
\end{align}
The sector with constant $\rho$ and $\sigma$ has broken dilatation symmetry and therefore is natural to use such covariant derivatives.

In table \ref{tablecc} we summarize the values of the coupling constants present in the action. 
\begin{table}
\begin{center}
\begin{tabular}{ll}
dual operator cases & range of validity \\ \hline
$\varepsilon_1=1$, $\varepsilon_2=1$ & \\
$\varepsilon_3=+1$ or $-1$ & for $N<4$ or $N>4$ respectively\\
$\varepsilon_s=+1$ or $-1$ & then $\rho>\sigma$ or $\rho<\sigma$ respectively\\
\\
coupling constant  & definition/ value \\ \hline
$\frac{1}{16\pi G_{N}}=\frac{M_p^2}{2}$ & $\varepsilon_s(\rho^2-\sigma^2)=\frac{\MP^2}{2\xi}$\\
$\Lambda$ & $\Lambda=-\varepsilon_s\frac{3M_p^2}{2\xi}$\\
$g^{(SU(N))}$ & $g^{(SU(N))}=\frac{1}{\sqrt{\xi}}$\\
$g^{(U(1))}$ & $g^{(U(1))}=\text{sign}(4/N-1)\left(\frac{\varepsilon_3}{8\xi}\left(\frac{4}{N}-1\right)\right)^{1/2}$\\
\\
$T_\text{NMC}$, $\calO(5)$ &  \\ \hline
$m_\text{eff}$ & $\sim\frac{\MP \cosh 2\lambda}{\sqrt{\xi}}$, see eqs. (\ref{fqt}); and (\ref{coshl1}) or (\ref{coshl2})\\
Pauli-like couplings & $\sim\frac{1}{\MP}F^r_{ab}\psibar\rho(\TT_r)\Sigma^{ab}\psi$ see eq. (\ref{quadraticfermion})\\
                     & $\sim\frac{\sqrt{\xi}}{\MP}\epsilon_{abcd}R^{abcd}$ see second line of eq. (\ref{quadraticfermion})\\
torsion coupling & $\sim T_{bcd}\epsilon^{abcd} \psibar \gf \gamma_a \psi$ see eq. (\ref{torsionterm_v2})\\
\\
$T_\text{NMC}$, $\calO(6)$ &  \\ \hline
$g_\text{fqt}$ & $\sim\frac{1}{\MP^2}$, see eq. (\ref{fqt})\\
\end{tabular}
\caption{Upper panel: defining parameters of the dual operator, see eq. (\ref{dualop}). Middle panel: gravity and gauge minimal couplings. Lower panel, dimension five operators: non minimal couplings present in the model. Parameters $\xi$ and $\lambda$ are phenomenological. The pauli-like gravitational coupling is proportional to the covariant derivative of the torsion 2-form, it vanish identically for a Riemannian connection. These terms may cancel exactly or not depending on the choice of $\varepsilon_s$, $\varepsilon_1$, $\varepsilon_2$, $\varepsilon_3$ and $\varepsilon_\psi$. The choice of such parameters also influences the susy conditions (\ref{intcondsusy}). Lower panel, dimension six operators: fermion quartic terms.}\label{tablecc}
\end{center}
\end{table}

Among the $T_\text{NMC}$ terms we have fermion quadratic terms that come from (\ref{LNMCpsi2}). After applying the scaling (\ref{chiralscaling}), and the scalings (\ref{globalscaling1}) and (\ref{globalscaling2}), we obtain the following phenomenological couplings
\begin{align}
 \calL(\psi^2)/(\dV)=&\frac{-i\sqrt{2}}{6\MP}\psibar  \ F^{\text{P-L}}_{ab} \ \Sigma^{ab}\psi\nonumber\\
 &-i\frac{(\varepsilon_\psi+\varepsilon_s)\sqrt{2\xi}}{24\MP}\epsilon_{ab}{}^{cd}R^{ab}{}_{cd}\psibar \psi\nonumber\\
 &-(\varepsilon_\psi-\varepsilon_s)\frac{\MP}{2\sqrt{2}}\frac{\cosh{2\lambda}}{\sqrt{\xi}}\psibar\psi\,.\label{quadraticfermion}
\end{align}
where
\begin{equation}
    F^{\text{P-L}}_{ab} = (\varepsilon_\psi-\varepsilon_2)F_{ab}^I\lambda_I+\sqrt{2}(\varepsilon_\psi-\varepsilon_3)\left(\frac{4/N-1}{\varepsilon_3}\right)^{1/2}F_{ab}-i\frac{(\varepsilon_\psi-\varepsilon_1)}{2}H_{ab}\gf\,,
\end{equation}
(these are all the final physical fields).
Second line in (\ref{quadraticfermion}) gets a $\sim \sqrt{\xi}$ scaling in the numerator thanks to the chiral scaling (\ref{chiralscaling}) of the scalar term prior to the global rescaling (\ref{globalscaling1}) and (\ref{globalscaling2}). This scaling comes from fact that the spin connection does not scale in the same way as the other gauge fields that are contained in $F^{\text{P-L}}_{ab}$.
The last term in (\ref{quadraticfermion}) is an effective mass term for the fermion. Note that many terms in (\ref{quadraticfermion}) may cancel exactly as a result of a given choice of the $\varepsilon$-parameters. In the last line of (\ref{quadraticfermion}) we used the parametrization
\begin{equation}\label{coshl1}
 \rho=\frac{1}{\sqrt{2\xi}}\cosh \lambda \ \MP\,,\quad \sigma=\frac{1}{\sqrt{2\xi}}\sinh \lambda \ \MP\,, \quad \text{for} \quad \varepsilon_s=+1\,,
\end{equation}
or
\begin{equation}\label{coshl2}
 \rho=\frac{1}{\sqrt{2\xi}}\sinh \lambda \ \MP\,,\quad \sigma=\frac{1}{\sqrt{2\xi}}\cosh \lambda \ \MP\,, \quad \text{for} \quad \varepsilon_s=-1\,.
\end{equation}
where $\lambda$ is an arbitrary parameter. Such parametrization produces the term $\sim \cosh 2\lambda /\sqrt{\xi}$ in the effective mass term of the third line in (\ref{quadraticfermion}) and it allows us to tune the value of the effective mass to a desired value. A relation emerges between $m_{eff}$ and $\Lambda$,
\begin{equation}
 m_{eff}^2 \ \sim \ |\Lambda| \cosh^2 2\lambda\,. 
\end{equation}
Therefore $m_\text{eff}$ acquires an extremely small value $m_\text{eff}\sim \Lambda^{1/2}$ unless a high level of tuning is invoked $\lambda \gtrsim 1$. A further phenomenological constant can be introduced by including an independent cosmological constant term in the action but we will not pursue this here.

Fermion quartic terms are also present in $T_\text{NMC}$.
\begin{equation}\label{fqt}
 \calL(\psi^4)/(\dV)= \frac{\left(\varepsilon_3\left(\frac{4}{N}-1\right) -\varepsilon_1\right)}{144 M_P^2}\phi^{ab}\phi_{ab}+\frac{\varepsilon_s}{12 M_P^2}(\phi^2+\phi^2_5)+\frac{\varepsilon_2}{36 M_P^2} \phi^{abI}\phi_{abI}\,,
\end{equation} 
There terms do not exhibit special cancellation for any particular choice of the $\varepsilon$-parameters, however they are highly suppressed and do not get contributions from the phenomenological parameter $\xi$.

The background torsion two form couples to the pseudo-scalar bilinear from terms in the first line in (\ref{LNMCpsi2}),
\begin{equation}\label{torsionterm_v2}
    T_\text{torsionNMC} = \frac{i}{6} T_{bcd}\epsilon^{abcd} \psibar \gf \gamma_a \psi\,,
\end{equation}
where $T^a = \half T^a{}_{bc}e^b e^c$. In the presence of fermion condensate, such a term allows to treat the completely antisymmetric component of the torsion two form as chemical potential. 

\section{Summary}\label{conclu}

In this paper we have defined a dual operator that allowed us to construct unified theories of gravity and matter. The appropriate definition of the dual operator allows us to:
\begin{enumerate}
 \item Prevent the appearance of ghosts in the $U(1)$ sector. Such technical difficulty is prevalent in models that gauge the full superconformal algebra.
 \item Use the tools of differential geometry in the study of the on-shell conditions of the broken symmetries, see eqs. (\ref{intcondexp}) and (\ref{intcondsusy}). Such analysis greatly simplifies the study of symmetries using the field transformations given in appendix \ref{symApp}. Such dual operator simplifies the field equations for the study of solutions.
 \item Introduce a natural notion of self duality in the context of this models, see eq. (\ref{selfdual}). 
\end{enumerate}

In section \ref{modelsummary} we gave details of the phenomenological parameters of the model in a sector of theory in which General Relativity is recovered. The promotion of $\rho$ and $\sigma$ to fields correspond to a sector of the theory with Weyl invariance (assuming for the moment the absence of conformal anomaly). Such sector of the theory correspond to a model where $\GN$ is not a fundamental scale, in the same spirit of Kaluza-Klein inspired solutions to the hierarchy problem where a Pati-Salam $SU(4) \times SU(2) \times SU(2)$ gauge symmetry is invoked in the bulk \cite{Antoniadis:1998ig,Arkani-Hamed:1998jmv}. We will study the phenomenology of the scalar modes $\rho$ and $\sigma$; and the embedding of the Pati-Salam or the $SU(5)$ GUT gauge groups in the present framework in a future work.

\section*{Acknowlegements}

We thank Cristobal Corral and Maria del Pilar Garcia for useful discussions. P. A. acknowledges MINEDUC-UA project ANT 1755 and Semillero de Investigación project SEM18-02 from Universidad de Antofagasta, Chile.

\begin{appendices}

\section{Fundamental representation of $SU(2,2|N)$}\label{representationApp}

Let us consider the following representation of $SU(2,2|N)$
\begin{eqnarray}
&\JJ_a =\left[\begin{array}{c|c}
\frac{s}{2}\gamma_a &  0_{4\times N}\\[0.5em] \hline
0_{N\times4} & 0_{N\times N} \\
\end{array}\right]\,, \quad \text{or} \quad (\JJ_a)^A_{\ B}=\frac{s}{2}(\gamma_a)^\alpha_{\ \beta}\delta^A_{\ \alpha} \delta^\beta_{\ B}=\frac{s}{2}(\gamma_a)^A_{\ B}\,,&\\
&\JJ_{ab} =\left[\begin{array}{c|c}
\frac{1}{4}[\gamma_a,\gamma_b] &  0_{4\times N }\\[0.5em] \hline
0_{N\times4} & 0_{N\times N} \\
\end{array}\right]\,, \quad \text{or} \quad (\JJ_{ab})^A_{\ B}=\frac{1}{4}[\gamma_a,\gamma_b]^A_{\ B}=(\Sigma_{ab})^A_{\ B}\,,&\\
&\KK_a =\left[\begin{array}{c|c}
\frac{1}{2}\tilde{\gamma}_a &  0_{4\times N}\\[0.5em] \hline
0_{N\times4} & 0_{N\times N} \\
\end{array}\right]\,, \quad \text{or} \quad (\KK_a)^A_{\ B}=\frac{1}{2}(\tilde{\gamma}_a)^A_{\ B}\,,&\\
&\DD =\left[\begin{array}{c|c}
\frac{1}{2}\gamma_5 &  0_{4\times N}\\[0.5em] \hline
0_{N \times4} & 0_{N\times N} \\
\end{array}\right]\,, \quad \text{or} \quad (\DD)^A_{\ B}=\frac{1}{2}(\gamma_5)^A_{\ B}\,,&\\
&\TT_{I} =\left[\begin{array}{c|c}
0_{4\times4} &  0_{4\times N}\\[0.5em] \hline
0_{N \times4} & \frac{i}{2}\lambda_{I}^{t} \\
\end{array}\right]\,, \quad \text{or} \quad (\TT_I)^A_{\ B}=\frac{i}{2}(\lambda_I^{t})_{\ B}^{A} \,,&\\
&(\QQ^\alpha_i)^A_{\ B}=\left[\begin{array}{c|c}
0_{4\times4} & 0_{4\times N}\\ [0.5em] \hline
\delta^A_{i} \delta^\alpha_B & 0_{N\times N}
\end{array}\right]=\delta^A_{i} \delta^\alpha_B\,,&\\
&(\QQb_\alpha^i)^A_{\ B}=\left[\begin{array}{c|c}
0_{4\times4} & \delta^A_\alpha \delta^i_B\\ [0.5em] \hline
0_{n\times 4} & 0_{N\times N}
\end{array}\right]=\delta^A_\alpha \delta^i_B\,,&\\
&\ZZ^A_{\ B}=z\left[\begin{array}{c|c}
i\delta^\alpha_\beta &0_{4\times N}\\ [0.5em] \hline
0_{n\times4} & \frac{4}{N} i\delta^i_j\end{array}\right]=
z\left(i\delta^A_\alpha\delta^\alpha_B+\frac{4i}{N}\delta^A_i\delta^i_B\right)\,,&
\end{eqnarray}
where $\gamma_5=i\gamma^0 \gamma^1 \gamma^2 \gamma^3$, $(\gamma_5)^2=\mathds{1}$,
$$\gamma_{abc}=\gamma_{[a}\gamma_b \gamma_{c]}=\frac{1}{3!}\sum_{\Pi(a,b,c)} \text{sign}(\Pi(a,b,c)) \gamma_a \gamma_b \gamma_c=i\epsilon_{abcd}\gamma_5 \gamma^{d}\,.$$
and $$\tilde{\gamma}_a=\frac{i}{3!}\epsilon_{abcd}\gamma^{bcd}=-\gamma_5\gamma_a\,,$$

The $\gamma$-matrices are in a $4\times 4$ spinor-representation ($\alpha, \beta,\cdots$ run from 1 to 4). The indexes of the tangent space $a,b=0,1,2,3$. Indexes in the adjoint representation of $su(N)$ take values $I,J=1,2,\ldots,N^{2}-1$, and in the fundamental take the values $i,j=1,2,\ldots,N$. The $\gamma$-matrices are endomorphisms and they act on spinors
\begin{equation}
 \psi^\alpha \stackrel{\gamma_a}{\longrightarrow} (\gamma_a)^\alpha_{\ \beta} \psi^\beta\,.
\end{equation}
These $\gamma$-matrices satisfy $\{\gamma^a,\gamma^b\}=2 \eta^{ab}$, where the metric $\eta$ is given by $\eta=\mathrm{diag}(-,+,+,+)$. The spinor indexes will be often omitted.

In a similar way the $\lambda$-matrices are also endomorphisms and they act on spinors as
\begin{equation}
 \psi^\alpha_i \stackrel{\lambda_I}{\longrightarrow} (\lambda_I)_i^{\ j}\psi^\alpha_j \,.
\end{equation}
The $\lambda$-matrices satisfy $[\lambda_I,\lambda_J]=f^{IJK} \lambda_K$, where indexes are raised/lowered with an Euclidean metric $\delta_{IJ}$. Indexes of the representation are $A,B=1,\cdots,N+4$, so we have a $N+4\times N+4$ representation. We find convenient to split $A=(\alpha,i)$. All the possible products that mix spaces like $p^i_{\ A}  q^A_{\ \alpha}$ are trivial. Thus, the following relations are understood
\begin{eqnarray}
&(\gamma_a)^A_{\ B}=\delta^A_\alpha (\gamma_a)^\alpha_{\ \beta} \delta^\beta_B\,,&\\
&C_{\alpha A}=C_{\alpha\beta} \delta^\beta_A \,,&
\end{eqnarray}

The generators $\JJ_a$ and $\JJ_{ab}$ form a adS$_4$ algebra,
\begin{equation}
[\JJ_a,\JJ_b]=s^2\JJ_{ab}\,,
\end{equation}
\begin{equation}
[\JJ_a,\JJ_{bc}]=\eta_{ab}\JJ_c-\eta_{ac}\JJ_b\,,
\end{equation}
\begin{equation}
[\JJ_{ab},\JJ_{cd}]=-(\eta_{ac}\JJ_{bd}-\eta_{ad}\JJ_{bc}-\eta_{bc}\JJ_{ad}+\eta_{bd}\JJ_{ac})\,.
\end{equation}
The parameter $s^2$ can take values $s^2=+1,-1$ for Anti de Sitter or de Sitter algebras respectively.

Among $\DD$ and $\KK_a$ they form the conformal algebra,
\begin{equation}
[\KK_a,\KK_b]=-\JJ_{ab}\,.
\end{equation}
\begin{equation}
[\JJ_a,\KK_{b}]=s\eta_{ab}\DD\,.
\end{equation}
\begin{equation}
[\KK_a,\JJ_{bc}]=\eta_{ab}\KK_c-\eta_{ac}\KK_b\,.
\end{equation}
\begin{equation}
[\DD,\KK_{a}]=-s^{-1}\JJ_a\,.
\end{equation}
\begin{equation}
[\DD,\JJ_{a}]=-s\KK_a\,.
\end{equation}

For the internal generators we have the $su(N)$ algebra
\begin{equation}
 [\TT_I,\TT_J]=f_{IJ}{}^K\TT_K\,,
\end{equation}
and they are anti-hermitian $\TT_I^\dag=-\TT_I$ (also $\ZZ^\dag=-\ZZ$).

Inlcuding $\QQ^\alpha_i$ and $\QQb^i_\alpha$ the commutators close in a $su(2,2|N)$ superalgebra
\begin{align}
&[\JJ_a,\QQb_\alpha^i]=\frac{s}{2}\QQb_\beta^i(\gamma_a)^\beta_{\ \alpha}\,, \quad [\JJ_a,\QQ^\alpha_i]=-\frac{s}{2}(\gamma_a)^\alpha_{\ \beta}\QQ^\beta_i\,,\\
&[\JJ_{ab},\QQb_\alpha^i]=\QQb_\beta^i(\Sigma_{ab})^\beta_{\ \alpha}\,, \quad [\JJ_{ab},\QQ^\alpha_i]=-(\Sigma_{ab})^\alpha_{\ \beta}\QQ^\beta_i\,,\\
&[\KK_a,\QQb_\alpha^i]=\frac{1}{2}\QQb_\beta^i(\tilde{\gamma}_a)^\beta_{\ \alpha}\,, \quad [\KK_a,\QQ^\alpha_i]=-\frac{1}{2}(\tilde{\gamma}_a)^\alpha_{\ \beta}\QQ^\beta_i\,,\\
&[\DD,\QQb_\alpha^i]=\frac{1}{2}\QQb_\beta^i(\gamma_5)^\beta_{\ \alpha}\,, \quad [\DD,\QQ^\alpha_i]=-\frac{1}{2}(\gamma_5)^\alpha_{\ \beta}\QQ^\beta_i\,,\\
&[\TT_I,\QQb_\alpha^i]=-\frac{i}{2}\QQb_\alpha^j (\lambda_I)^{\ i}_j\,, \quad [\TT_I,\QQ^\alpha_i]=\frac{i}{2}(\lambda_I)^{\ j}_i\QQ^\alpha_j\,,\\
&[\ZZ,\QQb_\alpha^i]=-iz(4/N-1)\QQb_\alpha^i\,, \quad [\ZZ,\QQ^\alpha_i]=iz(4/N-1)\QQ^\alpha_i\,,\\
&\{\QQ^\alpha_i,\QQb_\beta^j\}=\left(\frac{1}{2s}(\gamma^a)^\alpha_{\ \beta} \JJ_a-\frac{1}{2}(\Sigma^{ab})^\alpha_{\ \beta} \JJ_{ab}-\frac{1}{2}(\tilde{\gamma}^a)^\alpha_{\ \beta} \KK_a+\frac{1}{2}(\gamma_5)^\alpha_{\ \beta} \DD\right)\delta^j_i\nonumber\\
&\qquad \qquad \qquad +\delta^\alpha_\beta\left(-i(\lambda_I)_i^{\ j}\TT_I-\frac{i}{4z}\delta_i^{j} \ZZ\right)\,.
\end{align}

\subsection*{Traces:}
The graduation operator is given by
\begin{equation}
\mathcal{G}^A_{\ B} = \delta^A_{\alpha}\delta^\alpha_{\ B}-\delta^{A}_{i}\delta^i_{\ B}\,,
\end{equation}
it classifies generators in bosonic $B=\{\JJ_a,\JJ_{ab},\KK_a,\DD,\TT_I,\ZZ\}$ or fermionic $F=\{Q^\alpha_i,\QQb^i_\alpha\}$, by $[B,\mathcal{G}]=0=\{F,\mathcal{G}\}$, and it squares to one $\mathcal{G}^2=1$. With the graduation operator we can define an invariant supertrace
\begin{equation}
\langle G\rangle  \equiv Tr(\mathcal{G} G)=0\,.
\end{equation}
The supertrace have the following properties
\begin{equation}
 \langle B_1 B_2\rangle =\langle B_2 B_1\rangle \,, \quad \langle B F\rangle =\langle F B \rangle \,, \quad \langle F_1 F_2\rangle =-\langle F_2 F_1\rangle \,.
\end{equation}

All the generators $G$ in the representation are s-traceless
\begin{equation}
\langle G\rangle=0\,, \quad G=\{\JJ_a,\JJ_{ab},\KK_a,\DD,\TT_I,\ZZ,Q^\alpha_i,\QQb^i_\alpha\}\,.
\end{equation}
The quadratic combinations that give nontrivial traces are
\begin{eqnarray}
&\langle \JJ_a \JJ_b \rangle=s^2\eta_{ab}\,, \qquad \langle \JJ_{ab} \JJ_{cd} \rangle=-(\eta_{ac}\eta_{bd}-\eta_{bc}\eta_{ad})\,,&\label{trace1}\\
& \langle \KK_a \KK_b \rangle=-\eta_{ab}\,, \qquad \langle \DD^2\rangle = +1\,,&\\
&\langle \TT_I \TT_J \rangle=\frac{1}{2}\delta_{IJ}\,, \qquad \langle \ZZ^2\rangle=4z^2(4/N-1)\,,&\\
&\langle \QQ^\alpha_i \QQb^j_\beta\rangle=-\delta^\alpha_\beta \delta^j_i=-\langle \QQb^j_\beta \QQ^\alpha_i\rangle\,,&\label{trace-1}
\end{eqnarray}
and

\subsection*{$S$-grading operator}

The $S$ operator is fundamental in getting the usual expressions for the kinetic terms, the nontrivial traces are

\begin{align}
 \langle S \DD \rangle &=2i\varepsilon_s\,, \\
 \langle S \JJ_{ab}\JJ_{cd}\rangle&=-\varepsilon_s\epsilon_{abcd}=\langle \JJ_{ab}S\JJ_{cd}\rangle\,,\\
 \langle  \JJ_{a}S\KK_{b}\rangle&=-i\varepsilon_s s \eta_{ab}=-\langle \KK_{a}S\JJ_{b}\rangle\,,\\
 \langle \ZZ S \DD \rangle&=-2z\varepsilon_s=\langle \DD S \ZZ \rangle\,.
\end{align}

\section{$SU(2,2|N)$ curvatures}\label{curvaturesApp}

Here we give the explicity expresions of the curvatures.

The field strength is defined by
\begin{equation}
\FF=\half \calF^{ab} \JJ_{ab}+\calF^a \JJ_a +\calG^a \KK_a +\calH \DD +\calF^I \TT_I +\calF \ZZ +\QQb_\alpha^i \calX_i^\alpha +\ocalX_\alpha^i \QQ_i^\alpha\,,
\end{equation}
\begin{align}
\calF^{ab}&=\mathcal{R}^{ab}-\overline{\psi}^i\se\Sigma^{ab}\se\psi_i\,\\
\calF^a&=Df^a+\frac{1}{s}g^a h+\frac{1}{2s}\overline{\psi}^i\se\gamma^a\se\psi_i\,,\\
\calG^a&=Dg^a+sf^ah-\half \overline{\psi}^i\se\tilde{\gamma}^a\se\psi_i\,,\\
\calH  &=H+sf^ag_a+\half \overline{\psi}^i\se\gf \se\psi_i\,,\\
\calF^I&= F^I-i\overline{\psi}^i\se(\lambda^I)_i^{\ j}\se\psi_j\,,\\
\calF &=F-\frac{i}{4 z}\overline{\psi}^i\se\se\psi_i\,,\\
\calX_i^\alpha &=D(\se \psi_i)^\alpha+\frac{s}{2}f^a (\gamma_a\se \psi_i)^\alpha +\half g^a(\tilde{\gamma}_a\se \psi_i)^\alpha+\half h (\gf \se\psi_i)^\alpha\,,\\
\ocalX^i_\alpha&=-(\overline{\psi}^i\se)_\alpha\overleftarrow{D}+\frac{s}{2} (\overline{\psi}^i\se\gamma_a)_\alpha f^a +\half (\overline{\psi}^i\se\tilde{\gamma}_a)_\alpha g^a+\half (\overline{\psi}^i\se\gf )_\alpha h\,,
\end{align}
where
\begin{align}
H&=dh\,,\\
\mathcal{R}^{ab}&=R^{ab}+s^2f^af^b-g^ag^b\,,\\
 R^{ab}&=d\omega^{ab}+\omega^a_{\ c}\omega^{cb}\,,\\
 F^I&=dA^I+\frac{1}{2} f_{JK}^{I}A^J A^K\,,\\
 F&=dA\,,\\
 D_{\text{(Lorentz)-vector}}V^a&=dV^a+\omega^a_{\ b}V^b\\
 D_{\text{spinor}}\psi^\alpha&=d\psi^\alpha+\half \omega^{ab}(\Sigma_{ab}\psi)^\alpha-\frac{i}{2}A^I(\lambda_I\psi)_i^\alpha-i z(4/N-1)A\psi_i^\alpha\,,
\end{align}

The covariant derivative $D$ is defined for the $SO(1,3)\times SU(N)\times U(1)$ connection. The left-acting exterior derivative satisfies $\Omega^m\overleftarrow{d}=(-1)^m d\Omega^m$ for an $m$-form, in the spinor representation we get relations (\ref{covD}).

\section{Bianchi Identities}

Splitting along $S$-grading even, odd and fermionic parts we obtain
\begin{align}
 D\FF=&D_\Omega\FF^+ + D_\Omega\FF^-+[\Psi,\XX]+D_\Omega\XX-[\FF^++\FF^-,\Psi]\,,
\end{align}
The component expansion of $D\FF \equiv 0$ (along $\JJ_{ab}$, $\JJ_a$, $\KK_a$, $\DD$, $\TT_I$, $\ZZ$, and  $\QQb$ respectively) give us
\begin{align}
0\equiv& D_\omega \calF^{ab} + f^{[a}\calF^{b]}- g^{[a}\calG^{b]}-\ocalX \left(- \Sigma^{ab}\right)\se \psi +\psibar \se \left(-\Sigma^{ab}\right)\calX\,,\\
0\equiv&D_\omega \calF^a - \calF^a{}_b f^b- \calG^a h +\calH g^a-\ocalX \left(\half\gamma^a\right)\se \psi +\psibar \se \left(\half\gamma^a\right)\calX\,,\\
0\equiv&D_\omega \calG^a - \calF^a{}_b g^b- \calF^a h +\calH f^a-\ocalX \left(-\half\tgamma^a\right)\se \psi +\psibar \se \left(-\half\tgamma^a\right)\calX\,,\\
0\equiv&d \calH + \calG^a f_a- \calF^a g_a-\ocalX \left(\half \gf\right)\se \psi +\psibar \se \left(\half \gf\right)\calX\,,\\
0\equiv&D_{(A^J \TT_J)} \calF^I -\ocalX \left(-i\lambda_I\right)\se \psi +\psibar \se \left(-i\lambda_I\right)\calX\,,\\
0\equiv&d \calF -\ocalX \left(-\frac{i}{4z}\right)\se \psi +\psibar \se \left(-\frac{i}{4z}\right)\calX\,,\\
0\equiv&D_\Omega\calX  -\rho(\FF_\text{bos})\se\psi\,,
\end{align}
plus the complex conjugate of the last expression times $\QQ$, where
\begin{equation}
 \rho(\FF_\text{bos})=\half\calF^{ab}\Sigma_{ab}+\half \calF^a\gamma_a+\half \calG^a\tgamma_a+\half\gf \calH-\frac{i}{2}\lambda_I \calF^I-iz(4/N-1) \calF\,.
\end{equation}
In a purely bosonic background vacuum we have
\begin{align}
0\equiv& D_\omega \calR^{ab} + f^{[a}(D_\omega f^{b]}+g^{b]} h)- g^{[a}(D_\omega g^{b]}+f^{b]} h)\,,\\
0\equiv& D_\omega (D_\omega f^a+g^a h) - \calR^a{}_b f^b- (D_\omega g^a+f^a h) h +(H+f^b g_b) g^a\,,\\
0\equiv& D_\omega (D_\omega g^a+f^a h) - \calR^a{}_b g^b- (D_\omega f^a+g^a h) h +(H+ f^b g_b) f^a\,,\\
0\equiv& d (H+f^a g_a) + (D_\omega g^a+f^a h) f_a- (D_\omega f^a+g^a h) g_a\,,\\
0\equiv& D_{(A^J \TT_J)} F^I\,,\\
0\equiv& d F \,,
\end{align}
that greatly simplifies to
\begin{align}
0\equiv& D_\omega R^{ab}\,,\\
0\equiv&D_\omega (D_\omega f^a) - R^a{}_b f^b\,,\\
0\equiv&D_\omega (D_\omega g^a) - R^a{}_b g^b\,,\\
0\equiv& d (H+f^a g_a) + (D_\omega g^a) f_a- (D_\omega f^a) g_a\,,\\
0\equiv& D_{(A^J \TT_J)} F^I\,,\\
0\equiv& dF\,.
\end{align}

\section{Symmetry transformations}\label{symApp}

The symmetry transformations acting of the field components can be computed by using
\begin{equation}\label{covG}
\delta \AAA = D_\AAA G\,,
\end{equation}
where
\begin{equation}
G=\half \lambda^{ab}+\JJ_{ab}+\rho^a\JJ_a+\sigma^a\KK_a+\tau \DD+\theta^I \TT_I +\theta \ZZ+\QQb \epsilon +\overline{\epsilon} \QQ\,.
\end{equation}

From (\ref{covG}) we obtain
\begin{align}
\delta \omega^{ab}=&D_\omega \lambda^{ab} + 2f^{[a}\rho^{b]}- 2g^{[a}\sigma^{b]}-\left(\epsilonbar \Sigma^{ab}\se \psi +\psibar \se \Sigma^{ab}\epsilon \right)\,,\\
\delta f^a=&D_\omega \rho^a - \lambda^a{}_b f^b + \tau g^a -\sigma^a h +\half\left(\epsilonbar \gamma^a\se \psi +\psibar \se \gamma^a\epsilon\right)\,,\\
\delta g^a=&D_\omega \sigma^a - \lambda^a{}_b g^b + \tau f^a -\rho^a h -\half\left(\epsilonbar \tgamma^a\se \psi +\psibar \se \tgamma^a\epsilon\right)\,,\\
\delta h=&d \tau + f^a \sigma_a - g^a \rho_a +\half\left(\epsilonbar \gf\se \psi +\psibar \se \gf\epsilon\right)\\
\delta A^I=&D_{(A^J \TT_J)} \theta^I -i\left(\epsilonbar\lambda^I \se \psi +\psibar\se\lambda^I\epsilon\right)\,,\\
\delta A=& d \theta -\frac{i}{4z} \left( \epsilonbar \se\psibar+\psibar \se \epsilon \right)\,,\\
\delta (\se\psi)=&D_\Omega\epsilon -\rho(G_\text{bos})\se\psi \,,\\
\delta (\psibar\se)=&-\epsilonbar \lD_\Omega +\psibar\se \rho(G_\text{bos})\,,
\end{align}
where
\begin{equation}
 \rho(G_\text{bos})=\half \lambda^{ab}\Sigma_{ab}+\half\rho^a\gamma_a+\half\sigma^a\tgamma_a+\half \tau \gf-\frac{i}{2}\theta^I\lambda_I-i\theta z(4/N-1)\,.
\end{equation}

\end{appendices}

\bibliographystyle{ieeetr}
\bibliography{paper.bib}

\end{document}